\documentclass[aps,pra,twocolumn,groupedaddress,showpacs,showkeys]{revtex4-1}
\usepackage[utf8]{inputenc}
\usepackage{amsmath}
\usepackage{braket}
\usepackage{graphicx}
\usepackage{pgfplots}
\usepackage{csquotes}
\usepackage{multirow}
\usepackage{dcolumn}
\usepackage[colorlinks=true, citecolor=blue]{hyperref}
\usepackage{hhline}
\usepackage{amssymb}
\usepackage[none]{hyphenat}
\begin{document}
\title{Experimental realization of quantum cheque using a five-qubit quantum computer}

\author{Bikash K. Behera}
 \email{bkb13ms061@iiserkol.ac.in}
\affiliation{%
Indian Institute of Science Education and Research Kolkata, Mohanpur 741246, West Bengal, India
}%
\author{Anindita Banerjee}
\email{aninditabanerjee.physics@gmail.com}
\affiliation{%
 Department of Physics and Center for Astroparticle Physics and Space Science,
Bose Institute, Block EN, Sector V, Kolkata 700091, India
}%
\author{Prasanta K. Panigrahi}
\email{pprasanta@iiserkol.ac.in}
\affiliation{Indian Institute of Science Education and Research Kolkata, Mohanpur 741246, West Bengal, India}

\begin{abstract}
We demonstrate the implementation of quantum cheque, proposed by Roy Moulick and Panigrahi [Quantum Inf. Process (2016) 15: 2475], using the five-qubit IBM quantum computer. Appropriate single qubit, CNOT and Fredkin gates have been implemented for the realization of the quantum cheque transaction in a quantum networked banking system. \emph{Quantum state tomography} reveals the accuracy of the implementation with comparable results from the theoretical and experimental density matrices.     
\end{abstract}
\keywords{IBM Quantum Experience, Quantum Cheque, Quantum State Tomography}
\maketitle

\section{Introduction}
   
We make use of the free web based interface, \emph{IBM Quantum Experience} (IBM QE) \cite{IBM}, to experimentally demonstrate the quantum cheque transaction, proposed by Roy Moulick and Panigrahi \cite{SRM1}. Python \emph{Application Programming Interface} (API) and \emph{Software Development Kit} (SDK) \cite{QX17x}, have enabled easy writing of codes and running them on quantum processors. With fast access to the results of an experiment, the IBM QE users can communicate and discuss results with IBM researchers and other users. IBM QE permits a user easy-connectivity to this cloud \cite{IBM1} based 5-qubit quantum computer, using which a number of quantum algorithms \cite{QA1, QA2, QA3, QA4} and quantum computational tasks \cite{Devitt} have already been performed. Test of Leggett-Garg \cite{Huffman} and Mermin inequality \cite{Alsina}, quantum teleportation of an unknown single qubit \cite{Fedortchenko} and two qubit state \cite{Anirban0} have been reported. Entanglement assisted invariance \cite{Deffner}, non-Abelian braiding of surface code deffects \cite{Wootton}, and entropic uncertainty and measurement reversibility \cite{Berta} have been illustrated. A comparison between two architectures for quantum computation \cite{Linke} and non-destructive discrimination of Bell states \cite{Anirban} have also been experimentally performed. Here, we explicate experimental realization of quantum cheque transaction by implementing the scheme on IBM interface and find the accuracy of quantum state preparation through quantum state tomography.

Establishing long distance quantum communication networks \cite{Mor, Duan} is an active area of research, where a quantum cheque scheme \cite{SRM1} can be potentially used as an alternate for \emph{e-Payment Gateways} in the field of \emph{e-commerce}. It can also be considered as the quantum analog of the process of \emph{Electronic Data Interchange} (EDI) \cite{EDI}. Recently, practical unforgeable quantum money has been experimentally verified \cite{Nori,Diamanti}. The experimental demonstration therefore paves the way for designing of the physical devices for this purpose. In this scheme, efficient transactions can be performed by storing quantum states in computers or smart cards, equipped with quantum memories \cite{Trugenberger, Simon}. However, without quantum memory, the transactions can be streamed over the quantum internet \cite{Nemoto, Kimble} or the protocol can be run in real time, with the account holder physically going to the Bank, collecting a quantum cheque book, and then preparing a quantum cheque and issuing to a vendor. The vendor communicates the quantum cheque to the Bank and withdraws money after a successful verification of the cheque.

The paper is organized as follows. Section \ref{II} describes the implementation of quantum gates e.g., CNOT and Fredkin gates in order to design quantum circuits for experimental realization of quantum cheque. Section \ref{III} explicates the concept of a quantum cheque, following which, the implementation is shown on IBM Interface. Section \ref{V} demonstrates the accuracy of implementation by performing quantum state tomography. Finally, in section \ref{VI}, we conclude by summarizing our work and pointing out the future direction for further work.   

\section{Designing gates and some protocols on IBM QE \label{II}}
  
\begin{figure}[h]
    \centering
    \includegraphics[scale=0.5]{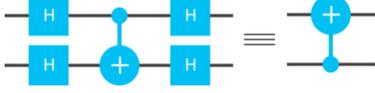}
    \caption{\textbf{Protocol-I}. \emph{Two equivalent quantum circuits showing implementation of CNOT gate in any order between two qubits on IBM interface.}}
     \label{figI}
\end{figure}

\begin{figure}[h]
     \centering
    \includegraphics[scale=0.5]{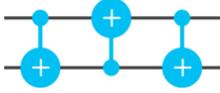}
    \caption{\textbf{Protocol-II}. \emph{Positioning of three CNOT gates for qubit swapping.}}
    \label{figII}
\end{figure}

For the implementation of a quantum cheque, one requires Hadamard (H), CNOT gate, the Pauli gates (X, Y and Z) and phase gates (S, $S^{\dagger}$, T and $T^{\dagger}$). Combining some of these gates, a fredkin gate can be constructed, which is used for the verification purposes of a quantum cheque. It is to be noted that, CNOT gate is not accessible to all five qubits on the interface of IBM, because of certain restrictions on the qubits. Protocol-I, depicted in Figure \ref{figI}, is used to construct CNOT gate in any order between two qubits. Similarly, Protocol-II, depicted in Figure \ref{figII}, is used to swap any two qubits on IBM interface. 

\section{Quantum Cheque \label{III}} 
A quantum cheque scheme is composed of three algorithms, Gen, Sign and Verify. Gen algorithm produces a \enquote{cheque book} and a key for the customer, who issues a cheque. Sign algorithm creates a quantum cheque state, $QC$, and Verify algorithm checks the validity of a cheque. A quantum cheque has mainly three properties, Verifiability, i.e., it can be verified by a Bank's main branch or any of its acting branches, Non-repudiation, i.e, after issuing a cheque, a customer must not be able to disclaim it, and Unforgeability, i.e., a quantum cheque can not be fabricated or it can not be reused. 

\subsection{The Quantum Cheque Scheme}
The quantum cheque scheme can be described by considering three parties, Alice, Bob and Bank. Here, the Bank is denoted as the main branch, which can have several branches securely connected by a classical channel. In this protocol, only Alice and Bank are considered to be trusted parties, and not necessarily the branches. After a cheque is issued by Alice, Bob goes to Bank or any of its branches to withdraw money.

The following three schemes are followed for a successful quantum cheque transaction.

\textbf{1. Gen Algorithm:} Initially, a shared key, $k$, is prepared by Alice and the Bank. Then Alice gives her public key, $pk$, to the Bank and collects her Private Key, $sk$.
 
The Bank prepares a set of $m$ number of GHZ states,

\begin{eqnarray}
\Ket{\phi^{(i)}}_{GHZ} = 	&\frac{1}{{\sqrt{2}}}    &   \big( \Ket{0^{(i)}}_{A_1}\Ket{0^{(i)}}_{A_2}\Ket{0^{(i)}}_{B} 	   \\ 	\nonumber
					&				& 	+ \Ket{1^{(i)}}_{A_1}\Ket{1^{(i)}}_{A_2}\Ket{1^{(i)}}_{B} \big)   \\ \nonumber
\end{eqnarray}

where $1 \leq i \leq m$, along with the respective unique serial number $s \in \{0,1\}^n$. From every GHZ entangled state, the Bank gives two qubits, named $|\phi\rangle_{A_{1}}$ and $|\phi\rangle_{A_{2}}$, and the serial number to Alice, while keeping the third qubit, $|\phi\rangle_{B}$, and other information, secretly in a database. 

Here, $ \{ |\phi^{(i)}\rangle_{GHZ}  \}_{i=1:m}$ stands for $ \{ |\phi^{(1)}\rangle_{GHZ}, |\phi^{(2)}\rangle_{GHZ}, \ldots, |\phi^{(m)}\rangle_{GHZ} \}$. 

At the end, Alice possesses $(id, pk, sk, k, s,  \{|\phi^{(i)}\rangle_{A_1}, |\phi^{(i)}\rangle_{A_2} \}_{i=1:m})$,
and the Bank carries $(id, pk, k, s, \{|\phi^{(i)}\rangle_{B} \}_{i=1:m})$. 

\textbf{2. Sign Algorithm:}
Alice prepares a random number by using a random number generation procedure, $r \leftarrow U_{\{0,1\}^L}$ to \emph{sign} a cheque of amount $M$ and creates a $n$-qubit state by using the following one way function \cite{crypt}, 
    $$|\psi_{alice}\rangle = f(k || id || r || M),$$ 
    
where, $k$ and $id$, are respectively the secret key and the identity of Alice. The symbol `$||$' concatenates two bit strings.

Alice also prepares $m$ states $\{|\psi_M^{(i)} \rangle \}_{i=1:m}$ corresponding to the amount M, using the one way function $g: \{0,1\}^* \times |0\rangle \rightarrow |\psi\rangle$, as $$ |\psi_M^{(i)}\rangle = g(r || M || i). $$ 

\begin{figure}[h]
    \centering
    \includegraphics[scale=1.0]{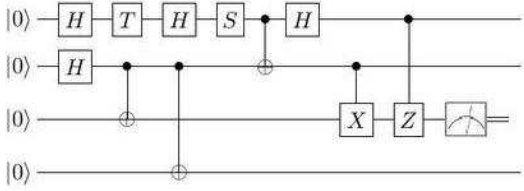}
    \caption{\emph{Depicting the quantum circuit used to generate a quantum cheque state.}}
    \label{fig:III}
\end{figure}

Subsequently, Alice encodes \cite{HBB99} $|\psi_M^{(i)} \rangle$ with the entangled qubit, $|\phi^{(i)}\rangle_{A_1}$ after which she performs a Bell measurement on her first two qubits as shown in Figure \ref{fig:III}.  

The state of the four qubit entangled system can be written in the following form,

\begin{eqnarray}
\label{eq:1}
	\begin{split}
	\Ket{\phi^{(i)}} =		& 	  \Ket{\psi_M^{(i)}} \otimes \Ket{\phi}_{GHZ} \\
			=		&	 \frac{1}{2} \big\{ \Ket{ \Psi^{+}}_{A_1} (\alpha_i \Ket{00}_{A_2B} + \beta_i \Ket{11}_{A_2B}) \\
					&	+ \Ket{\Psi^{-}}_{A_1} (\alpha_i \Ket{00}_{A_2B} - \beta_i \Ket{11}_{A_2B}) \\	
					&	+ \Ket{\Phi^{+}}_{A_1} (\beta_i \Ket{00}_{A_2B} + \alpha_i \Ket{11}_{A_2B}) \\  
					&	+ \Ket{\Phi^{-}}_{A_1} (\beta_i \Ket{00}_{A_2B} - \alpha_i \Ket{11}_{A_2B}) \big\}
	\end{split}
\end{eqnarray}

where $|\Psi^{+}\rangle, |\Psi^{-}\rangle, |\Phi^{+}\rangle$, and $|\Phi^{-}\rangle$ denote four Bell states. 

Now, Alice applies an appropriate Pauli gate operation on her qubit $|\phi^{(i)}\rangle_{A_2}$, according to the Bell state measurement outcomes:

\begin{center}
	\begin{tabular}{ l c*{4}    l }
		$|\Psi^{+}\rangle  \rightarrow I$ 			& & & & &	$|\Psi^{-}\rangle \rightarrow Z$ 	\\
		$| \Phi^{+}\rangle \rightarrow X$	& & & & &	$|\Phi^{-}\rangle \rightarrow Y$	\\
	\end{tabular}
\end{center}

Figure \ref{fig:III} depicts the encoding procedure of quantum cheque, which is to be repeated $m$ times.

Alice then uses sign algorithm to sign the serial number $s$ as $\sigma \leftarrow Sign_{sk}(s)$, and generates a quantum cheque $$ QC = (id, s,  r, \sigma, M, \{|\phi^{(i)}\rangle_{A_2}\}_{i=1:m}, |\psi_{alice}\rangle )$$ for Bob to encash.

\begin{figure}[h]
    \centering
    \includegraphics[scale=1.1]{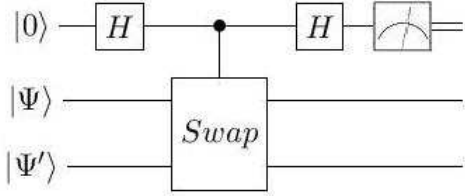}
    \caption{\emph{Quantum circuit for performing swap test on the two states $|\Psi\rangle$ and $|\Psi^{\prime}\rangle$.}}
    \label{fig:IV}
\end{figure}

\textbf{Swap Test:}
The swap test is depicted in Figure \ref{fig:IV}, where the measurement of ancilla (first qubit) on a computational basis yields zero if the two states $|\Psi\rangle$ and $|\Psi^{\prime}\rangle$ are equal. In this case, swap test is said to be successful. However, if the two states are different, then the measurement of ancilla yields both $|0\rangle$ and $|1\rangle$ each associated with some probability. For $\langle \Psi|\Psi^{\prime}\rangle$ $\geq$ $\lambda$, the swap test is successful with probability $\frac{1+\lambda^2}{2}$, and unsuccessful with probability $\frac{1-\lambda^2}{2}$. It is evident that, for the same input states, the swap test is successful with probability 1 and for different outputs, it is successful with probability less than 1. The efficiency of this test can be amplified by repeating it a large number of times. 

\textbf{3. Verify Algorithm:}
In the verification process, Bob produces the quantum cheque $QC$ at any of the acting branches of the Bank. The branch communicates with the Bank (main branch) to check the validity of the $(id,s)$ pair, and a verification is run by using $Vrfy_{pk}(\sigma, s)$. As described below, the Bank proceeds with the verification process if it finds $(id,s)$ and $\sigma$ to be valid, otherwise cancels the quantum cheque transaction. 

The Bank then measures its qubit, $|\phi_{B}\rangle$ in Hadamard basis to get $|+\rangle$ or $|-\rangle$ as output and conveys the results to the acting branch. The branch applies the appropriate Pauli gate operation on $|\phi^{(i)}\rangle_{A_2}$ to retrieve the unknown state $|\psi_M ^{(i)}\rangle $.

\begin{center}
	\begin{tabular}{ l c*{4}    l }
		$| +\rangle  \rightarrow I$ 			& & & & &	$|-\rangle \rightarrow Z$ 	\\
	\end{tabular}
\end{center}

A similar procedure is followed $m$ times to get $m$ unknown states $\{|\psi_M ^{(i)}\rangle \}_{i=1:m}$. The Bank generates $|\psi'_{alice}\rangle = f(k || id || r || M)$, and $\{|\psi_M^{,(i)} \rangle \}_{i=1:m} = \{g(r || M || i) \}_{i=1:m}$ by using these one way functions, and then performs a swap test on $m+1$ set of states, $\{|\psi_{alice}\rangle, |\psi'_{alice}\rangle\}$, and $\{|\psi_M^{(i)} \rangle, |\psi_M^{,(i)} \rangle \}_{i=1:m}$. 

The cheque is accepted if the swap test is successful, i.e., if $\langle \psi_{alice} \vert \psi'_{alice}\rangle \geq \lambda_1$ and $\{ \langle \psi_M^{(i)} \vert \psi_M^{,(i)}\rangle  \geq \lambda_2 \}_{i=1:m}$, where $\lambda_1$ and $\lambda_2$ are constants fixed by the Bank. Else, the branch terminates the transaction.

\section{Implementation of Quantum Cheque at IBM QE \label{IV}}

\begin{figure}[h]
    \centering
    \includegraphics[scale=0.17]{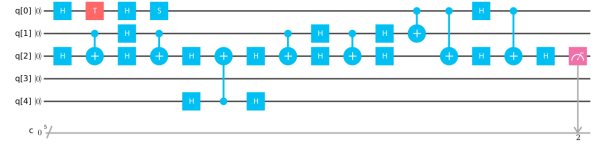}
    \caption{\emph{IBM quantum circuit used to generate the quantum cheque state.}}
    \label{fig:V}
\end{figure}

The IBM quantum circuit for generating a quantum cheque state has been depicted in Figure \ref{fig:V}. It is equivalent to the quantum circuit shown in Figure \ref{fig:III}. Though, these two figures appear to be different, their equivalency can be checked by using Protocol-I, Protocol-II and the concept of optimization of circuit. In Figure \ref{fig:V}, the first three qubits are in possession of Alice, the Bank contains the fifth qubit, and the fourth qubit remains unused. Alice uses one of her entangled qubit (2nd qubit), $|\phi^{(i)}\rangle_{A_1}$, provided by Bank, to encode the unknown state $|\psi_M^{(i)} \rangle$. Here, this unknown state can not be generated by using the one way function, ``g" (section \ref{III}), since we model only the quantum aspect and for simplicity only assume the g spits out a description of the following state, that is known to the preparation device but unknown to anybody else as \cite{Fedortchenko},  

\begin{equation} |\psi_M^{(i)}\rangle = cos(\pi/8)|0\rangle + sin(\pi/8)|1\rangle \end{equation} 

It can be computed by operating H, T, H and S gates sequentially on $|0\rangle$. As this state is now split between Alice's qubit (3rd qubit) and Bank's qubit (5th qubit), measuring the 3rd qubit in computational basis, it is expected to have $|0\rangle$ with probability $\approx$ 0.85 and $|1\rangle$ with probability $\approx$ 0.15. The experimental results are tabulated in Table \ref{tab1}. 

\begin{table}[h]
\centering
\caption{\emph{The table shows the results of the outcome of the quantum cheque state, depicted in Figure \ref{fig:V}, measured in computational basis. The results are obtained by both running and simulating the experiment with 1024, 4096 and 8192 number of shots.}}
\begin{tabular}{|c|c|c|}
\hline
\multicolumn{3}{|c|}{For Quantum Cheque Generation}\\
 \hline
 Number of Shots
 & Probability of $|0\rangle$ & Probability of $|1\rangle$\\
 \hline
Run-1 (1024) & 0.741 & 0.249\\
Run-2 (4096) & 0.766 & 0.234\\
Run-3 (8192) & 0.755 & 0.245\\ \hline
Simulation-1 (1024) & 0.848 & 0.152\\
Simulation-2 (4096) & 0.856 & 0.144\\
Simulation-3 (8192) & 0.856 & 0.144\\
 \hline
\end{tabular}
\label{tab1}
\end{table}

The encoding procedure (as described in section \ref{III}) should be done $m$ times by using $m$ similar quantum circuits (Figure \ref{fig:V}). Through the IBM cloud, it is not possible to create n-qubit quantum state by using a one way function, ``f" (section \ref{III}). So, we have taken two initial states $|0\rangle$, and $|0\rangle$ for comparison test.  

\begin{figure}[h]
    \centering
    \includegraphics[scale=0.2]{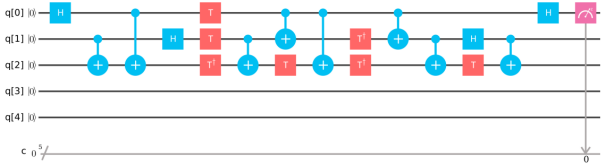}
    \caption{\emph{IBM quantum circuit used to verify the Quantum Cheque. It is to be noted that a set of two states are to be taken on the second and the third qubit of the above circuit for checking swap test.}}
    \label{fig:VI}
\end{figure}

The quantum circuits implemented on IBM interface, for quantum cheque verification, is illustrated in Figure \ref{fig:VI}, which is equivalent to the circuit shown in Figure \ref{fig:IV}. In this case, both the initial states (2nd qubit and 3rd qubit) are taken as, $|0\rangle$. (See Figure \ref{fig:VI}). It is expected to have $|0\rangle$ with probability 1, after measuring the ancilla qubit (1st qubit) in computational basis. The experimental results are illustrated in table \ref{tab2}. 

\begin{table}[h]
\centering
\caption{\emph{Table providing information about the ancilla state, depicted in Figure \ref{fig:VI}, when it is measured in computational basis. The experiment has been performed with 1024, 4096 and 8192 number of shots. Both run and simulated results are illustrated.}}
\begin{tabular}{|c|c|c|}
\hline
\multicolumn{3}{|c|}{For Quantum Cheque Verification}\\
 \hline
Number of Shots
& Probability of $|0\rangle$ & Probability of $|1\rangle$\\
 \hline
Run-1 (1024) & 0.813 & 0.188 \\
Run-2 (4096) & 0.839 & 0.161 \\
Run-3 (8192) & 0.846 & 0.154 \\ \hline
Simulation-1 (1024) & 1.000 & 0.000\\
Simulation-2 (4096) & 1.000 & 0.000\\
Simulation-3 (8192) & 1.000 & 0.000\\
 \hline
\end{tabular}
\label{tab2}
\end{table}

Comparing run result and simulated result, shown in tables \ref{tab1} and \ref{tab2}, run result is found to be less accurate than the simulated result. It is evident that application of a large number of gates increases decoherence of a quantum state and produces more noise in the system. Decoherence and noise due to gates are the key disadvantages for realizing the implementation of a quantum cheque with exact accuracy.  
 
\section{Quantum State Tomography \label{V}}
We now proceed to carry out state tomography to check how well, the quantum states are prepared in our experiment. We mainly consider two states, quantum cheque state ($|\phi^{(i)}\rangle_{A_2}$), which is to be stored in the quantum cheque, and ancilla state, used in swap test (See Section \ref{III}). In this process, by comparing both the theoretical and experimental density matrices of a quantum state, the accuracy of implementation can be tested. 

State tomography can be explained through a single qubit quantum state, $|\Psi\rangle = \alpha|0\rangle + \beta|1\rangle$. The theoretical and experimental density matrices of the given state are given by equations \ref{eq4} and \ref{eq5} respectively. 

\begin{equation}
\rho^{T}= |\Psi\rangle \langle\Psi|,     
\label{eq4}
\end{equation}
and 
\begin{equation}
\rho^{E}= \frac{1}{2} (I + \langle X\rangle X + \langle Y \rangle Y + \langle Z \rangle Z)
\label{eq5}
\end{equation}
 
Here, $\langle O \rangle = tr(|\Psi\rangle \langle \Psi|O)$, where O = X, Y, and Z. This expectation value can be obtained by rotating the quantum state along O axis and then measuring in computational basis. This can be evaluated as, $\langle O \rangle$ = P(0)-P(1), where P(0) and P(1) are the probabilities of outcome 0 and 1 respectively.  

The theoretical ($\rho^{T}_x,)$ and experimental ($\rho^{ER}_x, \rho^{ES}_x$) density matrices (both for run result and simulated result) of quantum cheque state ($x=q$) and ancilla state ($x=a$) are given below. 

\[
   \rho^{T}_{q}=
  \left[ {\begin{array}{cc}
   0.850 & 0.350 \\
   0.350 & 0.150 \\
  \end{array} } \right]
\]

\[
  \rho^{ER}_{q}=
  \left[ {\begin{array}{cc}
   0.760 & 0.043 \\
   0.043 & 0.240 \\
  \end{array} } \right]+ i\left[ {\begin{array}{cc}
   0.000 & -0.027 \\
   0.027 & 0.000 \\
  \end{array} } \right]
\]

\[
  \rho^{ES}_{q}=
  \left[ {\begin{array}{cc}
   0.852 & 0.008 \\
   0.008 & 0.148 \\
  \end{array} } \right]+ i\left[ {\begin{array}{cc}
   0.000 & -0.001 \\
   0.001 & 0.000 \\
  \end{array} } \right]
\]

\begin{figure}[h]
    \centering
    \includegraphics[width=\linewidth]{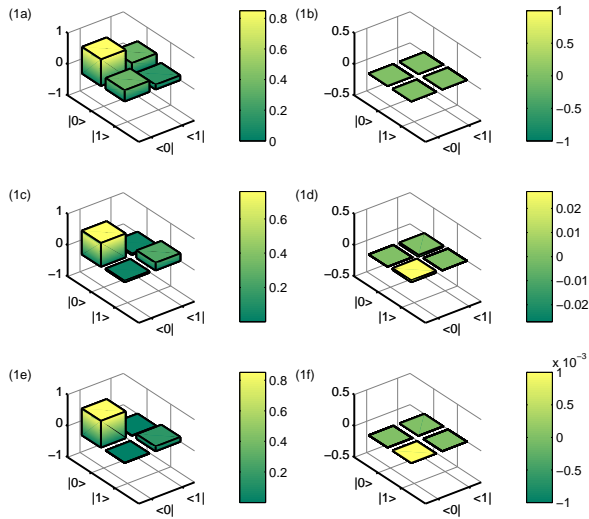}
    \caption{\textbf{Quantum cheque generation}: \emph{Real (left) and imaginary (right) parts of the reconstructed theoretical (1a,1b), run (1c,1d) and simulated (1e,1f) density matrices for the quantum cheque state.}}
    \label{fig:7}
\end{figure}

\[
   \rho^{T}_{a}=
  \left[ {\begin{array}{cc}
   1.000 & 0.000 \\
   0.000 & 0.000 \\
  \end{array} } \right]
\]
 
\[
   \rho^{ER}_{a}=
  \left[ {\begin{array}{cc}
   0.846 & 0.054 \\
   0.054 & 0.154 \\
  \end{array} } \right]+ i\left[ {\begin{array}{cc}
   0.000 & 0.062 \\
   -0.062 & 0.000 \\
  \end{array} } \right]
\]

\[
   \rho^{ES}_{a}=
  \left[ {\begin{array}{cc}
   1.000 & 0.009 \\
   0.009 & 0.000 \\
  \end{array} } \right]+ i\left[ {\begin{array}{cc}
   0.000 & -0.003 \\
   0.003 & 0.000 \\
  \end{array} } \right]
\]

\begin{figure}
    \centering
    \includegraphics[width=\linewidth]{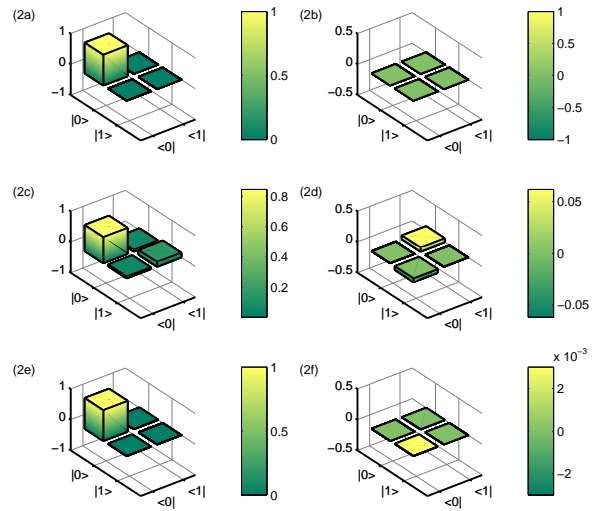}
    \caption{\textbf{Quantum cheque verification}: \emph{Real (left) and imaginary (right) parts of the reconstructed theoretical (2a,2b), run (2c,2d) and simulated (2e,2f) density matrices for the ancilla state.}}
    \label{fig:8}
\end{figure}

It is to be noted that the above experimental density matrices are calculated for running and simulating the experiment 8192 times. For other number of shots (1024 and 4096), similar procedure can be applied to obtain the corresponding density matrices. As mentioned in section \ref{IV}, by comparing the run, simulated and theoretical density matrices, it can be concluded that, the simulated result provides more accurate information, about the quantum state, as compared to the run results, which is already mentioned in section \ref{IV}. 

\section{Conclusion \label{VI}}
To conclude, here, we have demonstrated an experimental procedure of quantum cheque transaction in a quantum networked environment. Fredkin gate has been constructed, by using single qubit and CNOT gates, for verification of quantum cheque. The quantum state tomography has been performed to check the accuracy of the implementation. It is observed that the quantum cheque transaction has been carried out with good fidelity.         

\section*{Acknowledgments}
\label{acknowledgments}
B.K.B. is financially supported by DST Inspire Fellowship. We thank Subhayan Roy Moulick (Oxford) for contribution to the original concept of quantum cheque. We are extremely grateful to IBM team and IBM QE project. We thank IBM community for helpful suggestions regarding circuit construction. The discussions and opinions developed in this paper are only those of the authors and do not reflect the opinions of IBM or IBM QE team.

\end{document}